%% file: main_v4.tex
\documentclass[times,twocolumn,final]{elsarticle}
\usepackage{framed,multirow}
\usepackage{amsmath}
\usepackage{amssymb}
\usepackage{latexsym}
\usepackage{url}
\usepackage{xcolor}
\usepackage{hyperref}
\usepackage{booktabs}
\usepackage{color}
\usepackage{geometry}
\graphicspath{{./Figs/}}

\input{defs}

\definecolor{newcolor}{rgb}{.8,.349,.1}

\begin{document}
\begin{frontmatter} 
\title{Hybrid Representation Learning for Cognitive Diagnosis in Late-Life Depression Over 5 Years with Structural MRI\\
} 

\author{Lintao {Zhang}}
\author{Lihong {Wang}}
\author{Minhui {Yu}}
\author{Rong {Wu}}
\author{David C. {Steffens}}
\author{Guy G. {Potter}\corref{cor1}}
\author{Mingxia {Liu}\corref{cor1}}

\cortext[cor1]{Corresponding authors:  G.~Potter (guy.potter@duke.edu) and M.~Liu (mxliu@med.unc.edu).}

\begin{abstract}
Late-life depression (LLD) is a highly prevalent mood disorder occurring in older adults and is frequently accompanied by cognitive impairment (CI). 
Studies have shown that LLD may increase the risk of Alzheimer's disease (AD). However, the heterogeneity of presentation of geriatric depression suggests that multiple biological mechanisms may underlie it. 
Current biological research on LLD progression incorporates machine learning that combines neuroimaging data with clinical observations. 
There are few studies on incident cognitive diagnostic outcomes in LLD based on structural MRI (sMRI). 
In this paper, we describe the development of a hybrid representation learning (HRL) framework for predicting cognitive diagnosis over 5 years based on T1-weighted sMRI data. 
Specifically, we first extract prediction-oriented MRI features via a deep neural network, and then integrate them with handcrafted MRI features via a Transformer encoder for cognitive diagnosis prediction. 
Two tasks are investigated in this work, including (1) identifying cognitively normal subjects with LLD and never-depressed older healthy subjects, and (2) identifying LLD subjects who developed CI (or even AD) and those who stayed cognitively normal over five years. 
To the best of our knowledge, this is among the first attempts to study the complex heterogeneous progression of LLD based on task-oriented and handcrafted MRI features. 
We validate the proposed HRL on 294 subjects with T1-weighted MRIs from two clinically harmonized studies. 
Experimental results suggest that the HRL outperforms several classical machine learning and state-of-the-art deep learning methods in LLD identification and prediction tasks.  

\end{abstract}
\end{frontmatter}

\begin{figure*}[!t]
\setlength{\belowdisplayskip}{-1pt}
\setlength{\abovedisplayskip}{-1pt}
\setlength{\abovecaptionskip}{-1pt}
\setlength{\belowcaptionskip}{-1pt}
\centering
\includegraphics[width=1\textwidth]{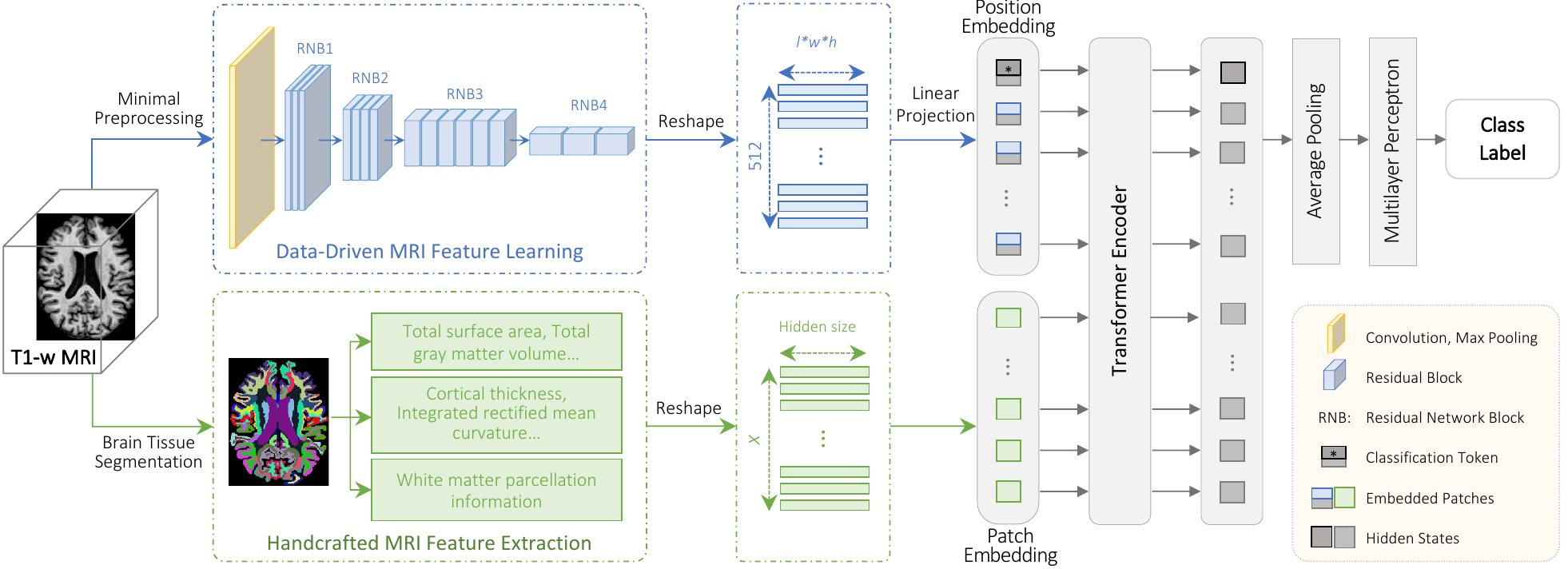}
\caption{Illustration of the proposed {\color{black}Hybrid Representation Learning (HRL)} framework for longitudinal diagnostic discrimination in LLD. The HRL consists of 4 components: (1) \emph{Data-Driven MRI Feature Learning} via a ResNet34 backbone, (2) \emph{Handcrafted MRI Feature Extraction} using tissue segmentation and parcellation tools, (3) \emph{Feature Fusion and
Abstraction} via Transformer Encoder, and (4) \emph{Classification}. The 512 MRI feature maps of size $l*w*h$ extracted by ResNet34 backbone are reshaped and projected to 512 embedded patches with uniform length (i.e., a parameter named hidden size in Transformer encoder) for position encoding. The handcrafted features are also reshaped to embedded patches of hidden size; the number of patches $x$ is depending on the length of the handcrafted feature vector. Then the embedded patches are concatenated as the input of the Transformer encoder. The attention mechanism of the Transformer encoder is utilized for better fusing and representing the data-driven MRI features and handcrafted MRI features.
}
\label{fig-network}
\end{figure*}
\section{Introduction}
\label{S1}
{L}ate-life depression (LLD), mild cognitive impairment (MCI), and dementia are prevalent disorders worldwide that affect older adults. 
Previous studies have shown a close relationship between depression, cognitive impairment (CI), and progressive dementia in late life, especially Alzheimer's disease (AD)~\citep{ly2021late, rashidi2020frontal, burke2019diagnosing, alexopoulos2019mechanisms, geerlings2017late, lebedeva2017mri, joko2016patterns}. 
Current research considers these entities related such that late onset LLD may be a prodromal symptom of dementia~\citep{burke2019diagnosing}. While there are multiple clinical pathways between these three disorders, the mechanism of action that links them is complex and poorly understood~\citep{butters2008pathways}; thus, the pathologic mechanisms remain unclear. 

Existing studies on the diagnosis of depression are mainly based on questionnaires and clinical interviews rather than using clinically relevant biomarkers
~\citep{burdisso2019text, abrams2019changes, balsamo2018assessment}. 
The lack of recognized objective biomarkers may lead to a high degree of diagnostic heterogeneity, which complicates the task of identifying etiologies and predicting outcomes ~\citep{hermida2012association}.
It is generally challenging to find objective biomarkers for predicting cognitive decline in LLD and including development of AD that might promote early intervention and effective treatment of the disease. 
Neuroimaging provides a promising method for understanding the complex pathophysiological progress of LLD that may support biomarker-driven diagnosis and treatment. 
Specifically, structural magnetic resonance imaging (sMRI) provides a non-invasive solution for objectively quantifying physical disorders that lead to significant mental illness. 
Increasing evidence has shown that local white matter and gray matter changes are directly related to depressive symptoms~\citep{guan2021cost, teodorczuk2010relationship, rashidi2020frontal}. 
Several studies have tried to discriminate LLD patients with differential cognitive progression based on sMRI~\citep{mousavian2019depression, lebedeva2017mri, joko2016patterns}. 
These MRI-based methods focus on classification, detection, and prediction of MCI, AD, and LLD with diagnosis information of baseline or 1-year follow up time. 
Few studies pay attention to longitudinal analysis of diagnostic cognitive change in LLD with sMRI.

To this end, we propose a Hybrid Representation Learning (\textbf{HRL}) framework for longitudinal diagnostic discrimination in LLD based on sMRI. 
The hypothesis is that the effective fusion of data-driven and handcrafted MRI features helps improve predictive performance of the deep learning model. 
As shown in Fig.~\ref{fig-network}, the HRL consists of 4 components: (1) data-driven MRI feature learning, (2) handcrafted MRI feature extraction, (3) feature fusion and abstraction, and (4) classification. 
We evaluate the HRL on 294 subjects from two studies, and the experimental results suggest its effectiveness in detection and prediction tasks related to diagnostic cognitive change in LLD. 
The source code has been released to the public via GitHub~\footnote{https://github.com/goodaycoder/LLDprogression}.

This paper is built upon our conference paper~\citep{zhang2022MLMI} with notable improvements. 
(1) Besides data-driven MRI features extracted by a deep neural network, we employ diverse handcrafted MRI features such as surface area, cortical thickness and gray matter volume.  
(2) Data-driven and handcrafted features of T1-weighted MRI are integrated into a unified framework through a Transformer encoder module. 
(3) We visualize the most informative brain regions that contributed to the prediction  task. These brain regions may contain potential biomarkers for diagnostic outcome prediction in LLD. 
(4) More experiments and ablation studies are conducted to demonstrate the effectiveness of each component of the proposed HRL. 
\if false
(1) In the previous work, FSL~\citep{jenkinson2012fsl} and ANTs~\citep{avants2009advanced} are used to process the MRI data and extract features of image intensity. We further extract more handcrafted features of brain structures from the MRI data with FreeSurfer~\citep{fischl2012FreeSurfer} in this work. 
(2) We combine the original classification and prediction model with a modified Transformer encoder from ViT~\citep{dosovitskiy2020image}, which can integrate handcrafted features and 3D MRI data into one framework to improve model prediction accuracy. 
(3) We conduct more experiments and ablation studies on the classification and prediction tasks to demonstrate the effectiveness of the improved approach. 
(4) More importantly, we apply our method to neuroimage analysis by visualizing and analyzing the most informative brain regions that contributed to the classification. These brain regions might contain the related potential biomarkers for LLD progression.  
\fi
To the best of our knowledge, this is among the first attempts dedicated to predicting cognitive diagnosis in LLD over 5 years and finding possible biomarkers from sMRI scans. 
The main contributions of this work are summarized as follows:
\vspace{-1mm}
\begin{itemize}
    \item A Hybrid Representation Learning (HRL) framework is developed for predicting diagnostic outcome of a 5-year longitudinal period in LLD using sMRI data. Compared to the MRI-based classification models in LLD-related studies, HRL integrates data-driven and handcrafted features of sMRI into a unified framework. 
    \vspace{-2mm}
    \item To identify structural MRI-based imaging biomarkers of the longitudinal diagnostic outcome, we visualize feature maps extracted by HRL and the most informative brain regions in diagnostic outcome prediction tasks. 
    \vspace{-2mm}
    \item Extensive experiments are conducted to validate the effectiveness of HRL in LLD identification, LLD-to-CI 
    and LLD-to-AD diagnostic outcome prediction.
\end{itemize}

The remainder of this paper is organized as follows. 
Section~\ref{S2} reviews the most relevant studies. 
In Section~\ref{S3}, we introduce the participants and proposed method. 
In Section~\ref{S4}, we compare our method with several competing methods for LLD identification and LLD-to-CI diagnostic outcome prediction, and analyze the most informative brain regions that might contain the related potential biomarkers for diagnostic outcome prediction in LLD. 
We further analyze several important aspects related to the performance of HRL, apply our HRL on LLD-to-AD diagnostic outcome prediction, and analyze the limitations of the current work in Section~\ref{S5}.  
This paper is finally concluded in Section~\ref{S6}.

\section{Related Work}
\label{S2}

\subsection{Studies on Late-Life Depression Progression}
In recent years, more attention has been drawn to late-life depression (LLD) progression. 
Butters et al. summarized and analyzed the possible pathways that link LLD to MCI and AD, and further proposed predominant mechanisms by which depression increases the risk for AD~\citep{butters2008pathways}. 
Plassman et al. included a pathway of persistent cognitive impairment that was neither clearly progressive nor clearly prodromal AD, which for in the context of our studies we have diagnosed as cognitive impairment, no dementia (CIND)~\citep{plassman2008prevalence}.
A 5-year longitudinal study using statistical analysis methods was conducted by Ly et al., and the results suggested that LLD participants of late-onset depression subtypes experienced faster cognitive decline than normal control (NC)~\citep{ly2021late}. 
In addition to statistical research based on clinical neuropsychological assessment, there were some LLD-related studies based on brain MRI data. 
Lebedeva et al. showed that LLD patients developing MCI and dementia could be discriminated from LLD patients remaining cognitively stable with good accuracy based on baseline structural MRI alone~\citep{lebedeva2017mri}. 
Mousavian et al. investigated machine learning algorithms for major depression disorder (MDD) detection from MRI Images ~\citep{mousavian2019depression}. 
A previous study in~\citep{joko2016patterns} suggested that the method of comparing hippocampal atrophy by region might be useful in distinguishing AD, MCI, MDD, and NC. 
These studies either used machine learning methods that rely on handcrafted features or used deep learning models for MRI-based LLD classification. 
Moreover, these tasks were based on immediate diagnosis at MRI scan time, and there is no relevant literature on long-term research of LLD progression.
In this work, we aim to perform a longitudinal study of predicting 5-year diagnostic outcome in LLD with sMRI.

\subsection{Brain MRI Representation Learning}
Previous studies have shown that sMRI could provide information for identifying depression~\citep{zhuo2019rise, binnewies2021associations}. 
In recent years, much research has been devoted to neuroimage analysis for computer-aided diagnosis of brain diseases~\citep{sarmento2019automatic, zhang2018multi}. 
Some studies tried to analyze potential biomarkers extracted from sMRI using machine learning~\citep{gao2018machine, nouretdinov2011machine}. 
For machine learning-based classification, the key issue is feature selection and reduction. 
Nouretdinov et al. suggested that feature selection determines the reliability of the predictions~\citep{nouretdinov2011machine}. 
Khan et al. used extreme learning machine to select deep learning features for better feature fusion and classification~\citep{khan2020multimodal}. 
Liu et al. suggested that the low-dimensional handcrafted volumetric features of the brain could preserve the biological information density and models trained with them yielded comparable performance to those utilizing whole-brain MRI~\citep{liu2019using}. 

\begin{figure}[!t]
\setlength{\belowdisplayskip}{-2pt}
\setlength{\abovedisplayskip}{-1pt}
\setlength{\abovecaptionskip}{-4pt}
\setlength{\belowcaptionskip}{-2pt}
\centering
\includegraphics[width=0.48\textwidth]{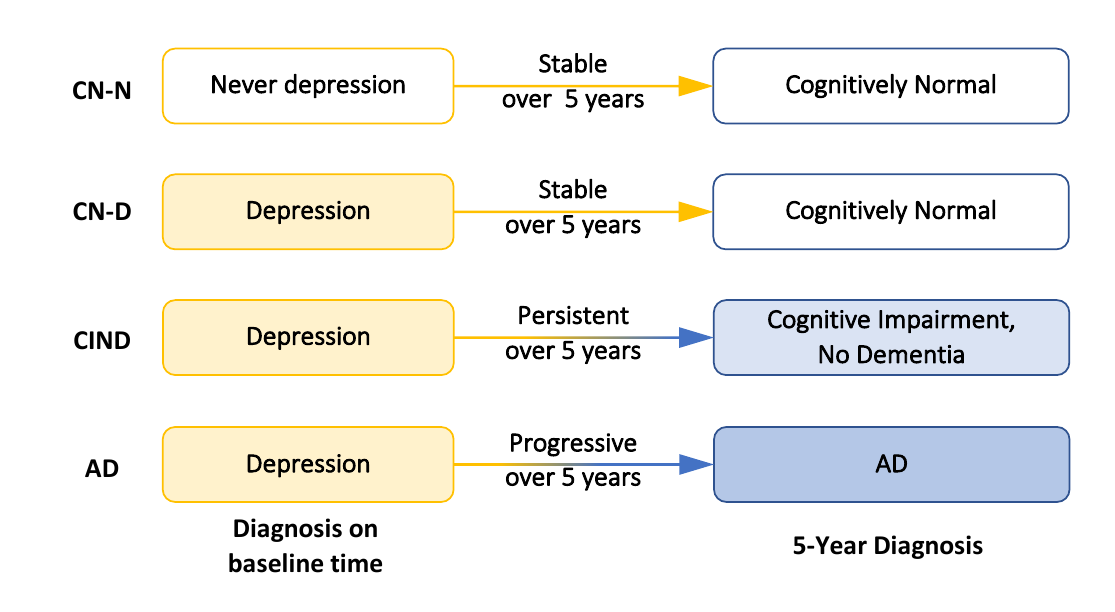}
\caption{Illustration of possible pathways of cognitive change in our study. The subjects of CIND and AD categories are combined into a CI group in the following classification task due to the limited sample size.}
\label{fig-pathway}
\end{figure}

Many deep learning methods have been proposed for data-driven MRI feature extraction and imaging biomarker exploration. 
For example, a degenerative adversarial neuroimage net was developed using sMRI data to capture the patterns of regional brain intensity changes that accompany disease progression ~\citep{ravi2019degenerative}. 
Ghazi et al. utilized a recurrent neural network (RNN) to model the progression of AD using six biomarkers of sMRI~\citep{ghazi2019training}.
Uyulan et al. built an electroencephalography-based diagnosis model for MDD diagnosis with CNN to explore translational biomarkers~\citep{uyulan2021major}. 
There are several studies that have tried to fuse the deep learning features with handcrafted features for 
disease diagnosis and tumor detection~\citep{shankar2021novel,bansal2022detection, hagerty2019deep} with 2D medical images. 
Saba et al. concatenated the deep learning features 
with handcrafted (shape and texture) features to achieve accurate and fast classification of brain tumors~\citep{saba2020brain}. 
However, effectively fusing deep learning features and handcrafted features is often challenging due to their inherent heterogeneity.

Transformer has recently been successfully applied to data fusion~\citep{shamshad2022transformers,zheng2022transformer,xing2022nestedformer}, thanks to its multi-head self-attention mechanism in selectively fusing image features.
Zheng et al. proposed a transformer-based multi-feature fusion model to fuse cortical features in MCI conversion prediction ~\citep{zheng2022transformer}. 
Xing et al. conducted multi-modal MRI feature fusion for brain tumor segmentation with a spatial attention transformer and a cross-modality attention transformer~\citep{xing2022nestedformer}.
These studies focused on handcrafted features or deep learning features separately. 
Inspired by these studies, in this work, we propose a hybrid representation learning framework with a Transformer encoder that can integrate deep learning and handcrafted MRI features for diagnostic outcome prediction.

\begin{table}[!tbp]
\renewcommand\arraystretch{1.1}
\centering
\setlength{\belowdisplayskip}{0pt}
\setlength{\abovedisplayskip}{0pt}
\setlength{\abovecaptionskip}{0pt}
\setlength{\belowcaptionskip}{0pt}
\scriptsize
\caption{Selection criteria of studied subjects in our work. The diagnoses are made in a 5-year follow-up time window from baseline.}

\setlength{\tabcolsep}{5mm}{}{
\begin{tabular}{l|l}   
\toprule
\multicolumn{1}{l}{Category} & \multicolumn{1}{|l}{Description} \\
\midrule 
CN-N &Cognitively Normal\\
\hline
CN-D &Depressed, Cognitively Normal \\
\hline
\multirow{2}{*}{CIND}   &Cognitive Impairment, No Dementia;\\
                        &Cognitive Impairment, due to Vascular Disease \\
\hline
\multirow{2}{*}{AD} &Probable AD, Possible AD, Subsyndromal AD;\\
                    &Dementias of undetermined etiology\\
\bottomrule
\end{tabular} 
\label{Diagnosis_information}
}
\end{table}

\section{Materials and Methodology}
\label{S3}
\subsection{Studied Subjects and Image Preprocessing}
\textbf{Participants and MRI Acquisition}. 
The studied subjects are enrolled in two studies: (1)  Neurocognitive Outcomes of Depression in the Elderly (NCODE) study~\citep{steffens2004methodology} and (2) Neurobiology of Late-life Depression (NBOLD) study~\citep{steffens2017negative}.
Both studies included individuals with LLD and a comparison sample of never-depressed controls. 
The 3T T1-weighted MRIs are included in this work. 
The MRI of NCODE was acquired under a 3 Tesla whole-body MRI Siemens Trio system (Siemens Medical Systems, Malvern, PA), and processed by the Neuro-psychiatric Imaging Research Laboratory (NIRL), located at Duke University Medical Center. 
The 3D axial TURBOFLASH sequence was used, with TR/TE=22/7~$ms$, flip angle=25°, a 100 Hz/pixel bandwidth, a 256×256 matrix, a 256~$mm$ diameter field-of-view, 160 slices with a 1~$mm$ slice thickness, and voxel size=1×1×1~$mm^3$. 
The MRI of NBOLD was acquired using a Skyra 3T scanner (Siemens, Erlangen, Germany) located at Olin Neuropsychiatric Research Center (ONRC), 
using a magnetization-prepared rapid gradient-echo (MPRAGE) protocol with TR/TE=2,200/2.88~$ms$, flip angle=13°, matrix=220×320×208, and voxel size=0.8×0.8×0.8~$mm^3$.

\textbf{Subject Selection and Grouping.}
Depression diagnoses are assigned by trained psychiatrists using standardized assessment instruments and diagnostic algorithms at enrollment time, as described in~\citep{steffens2004methodology, steffens2017negative}. 
All individuals included in this study were diagnosed as cognitively normal as the time of the baseline assessment. Importantly, subjects in both studies were diagnosed with the same instruments and protocol, assessed annually with the same battery of neuropsychological tests, and adjudicated for cognitive diagnosis by a group of experts following the same consensus diagnostic guidelines. Diagnosis at Year 5 in the study is the outcome of interest in this study, which included cognitively normal (CN), cognitive impairment , no dementia (CIND), or AD. At the follow-up visit time (i.e., 5 years), subjects stayed stable or converted to CIND, or AD are recorded, while the possible pathways included in this work are illustrated in Fig.~\ref{fig-pathway}.
The detailed subject selection criteria for this work are shown in Table~\ref{Diagnosis_information}. 
Following the criteria in Table~\ref{Diagnosis_information}, 311 subjects in total are initially selected for image preprocessing. After MRI preprocessing, 17 subjects are excluded due to the failed processing and low image quality. 
Finally, 294 subjects are included in our study, and the demographic information is shown in Table~\ref{Demographic_information}. 
Using MRI scan time as the baseline, category labels are given based on the 5-year diagnosis. 
As shown in Fig.~\ref{fig-pathway}, these subjects are categorized into 4 groups: 
(1) {87} never-depressed cognitively normal (CN-N) subjects: CN-N subjects at baseline who don't progress to cognitive impairment within the 5-year follow-up;
(2) {172} depressed CN (CN-D) subjects: CN-D subjects at baseline but cognitively normal at diagnosis and within the 5-year follow-up; 
(3) {19} CIND subjects: depressed subjects who develop cognitive impairment but not dementia in 5-year follow-up; 
(4) {16} AD subjects: depressed participants with a diagnosis of AD at 5 years after baseline.

\textbf{Prediction Task.} Table~\ref{Demographic_information} shows that each category has a limited number of subjects, especially the CIND and AD groups. 
So, we perform the diagnostic outcome prediction study in 
two classification tasks: (1) binary classification for LLD detection (i.e., CN-D vs. CN-N classification) and (2) three-category classification for predicting 5-year cognitive diagnosis (i.e., CI vs. CN-D vs. CN-N classification). 
In the task of three-category classification, CIND and AD subjects are combined into the CI group, which helps to increase the sample size. 
Besides, the CIND-to-AD diagnostic outcome prediction (i.e., CIND vs. AD classification) is also performed using a transfer learning strategy. 
That is, we first use 795 subjects selected from ADNI~\citep{jack2008alzheimer} for model training and then transfer the trained model to CIND and AD subjects in this study for testing, with details given in Section~\ref{S5.4}.

\begin{table}[!tbp]
\renewcommand\arraystretch{1.1}
\centering
\setlength{\belowdisplayskip}{0pt}
\setlength{\abovedisplayskip}{0pt}
\setlength{\abovecaptionskip}{0pt}
\setlength{\belowcaptionskip}{0pt}
\scriptsize
\caption{Demographic information of studied subjects in our work. The values are denoted as ``mean $\pm$ standard deviation''. F/M: Female/Male, MADRS: Montgomery-Asberg Depression Rating Scale~\citep{fantino2009self}, MMSE: Mini-mental state examination~\citep{galea2005mini}.}

\setlength{\tabcolsep}{4pt}{}{
\begin{tabular}{l|cc cc cc c}
\toprule
~{Dataset} &{Category} &{Gender (F/M)} & {Age} &{Education} & MADRS & MMSE\\
\midrule 
\multirow{4}{*}{NCODE}   & CN-N &31/22   &67.5$\pm$5.0 &15.7$\pm$1.4 &- &29.1$\pm$1.0\\
                         & CN-D & 67/39  &66.1$\pm$5.7 &15.0$\pm$2.2 &15.3$\pm$8.1 &28.5$\pm$1.4 \\
                         & CIND & 5/5    &71.9$\pm$6.3 &14.4$\pm$2.1 &16.7$\pm$8.8 &27.4$\pm$2.2 \\
                         & AD   & 5/3    &72.0$\pm$5.6  &14.4$\pm$2.1 &18.4$\pm$10.5 &24.8$\pm$5.4 \\
\hline
\multirow{4}{*}{NBOLD}   & CN-N & 27/7   &71.6$\pm$7.1 &15.6$\pm$2.1 &- &29.1$\pm$1.5 \\
                         & CN-D & 50/16  &70.0$\pm$6.9 &16.3$\pm$2.3 &19.7$\pm$5.5 &29.5$\pm$0.8 \\
                         & CIND & 6/3    &74.9$\pm$4.3 &14.3$\pm$2.7 &17.5$\pm$10.7 &29.2$\pm$1.2 \\
                         & AD   & 3/5    &76.5$\pm$7.5 &17.8$\pm$1.5 &17.5$\pm$7.1 &28.5$\pm$1.7 \\
                          
\bottomrule
\end{tabular} 
\label{Demographic_information}
}
\end{table}
\textbf{Image Preprocessing}.
Each sMRI scan is preprocessed using the FSL, ANTs, and FreeSurfer tools. 
The processing pipeline includes (1) bias field correction with N4, (2) linear registration to AAL3~\citep{rolls2020automated} template in the Montreal Neurological Institute (MNI) space, resampling to 1~$mm^3$ resolution, and cropping the whole brain to the size of 181×217×181~$mm^3$, (3) brain extraction, (4) non-linear registration to AAL3, and (5) partition regions-of-interest (ROI) of AAL3 into the registered sMRI volumes.
All MRI scans have the same size as the MNI template after preprocessing.

The average image intensities of gray matter (GM), white matter (WM), and cerebrospinal fluid (CSF) are computed within the 170 ROIs defined by AAL3. 
Nonlinear registration is performed for warping the ROIs of AAL3 back to each subject using ANTs.
{\color{black}Three 166-dimensional feature vectors 
are generated after eliminating 4 tiny ROIs for each subject, 
used as handcrafted features for the following classification task based on machine learning methods.}   
Furthermore, anatomical structural features are extracted using FreeSurfer, such as volume, cortical thickness, mean curvature, surface area, and white matter parcellation information, with more details given in Section~\ref{S3_2_2}.
{\color{black}
All MRIs are also preprocessed via histogram standardization, zero-mean normalization, and rescaling the intensity between [0, 1] using TorchIO~\citep{perez2021torchio}\footnote{https://torchio.readthedocs.io/}}.

\subsection{Proposed Method}
\if false
In this work, we perform a challenging classification task for assessing cognitive change in LLD. 
In order to combine the advantages of deep learning and handcrafted features, we proposed HRL framework that can take both the MRI data and handcrafted features as inputs.
\fi
We reasonably assume that effective fusion of data-driven and handcrafted MRI features can promote detection and prediction of diagnostic cognitive change in LLD, and  
exploit both types of MRI features via a hybrid representation learning (HRL) framework. 
As shown in Fig.~\ref{fig-network}, the HRL consists of 4 components: (1) {data-driven MRI feature learning} with a residual neural network (ResNet)~\citep{he2016deep} as backbone, (2) {handcrafted MRI feature extraction}, 
(3) {feature fusion and abstraction} 
via a Transformer encoder, and (4) {classification}. 

\begin{figure}[!t]
\setlength{\belowdisplayskip}{0pt}
\setlength{\abovedisplayskip}{0pt}
\setlength{\abovecaptionskip}{0pt}
\setlength{\belowcaptionskip}{0pt}
\centering
\includegraphics[width=0.49\textwidth]{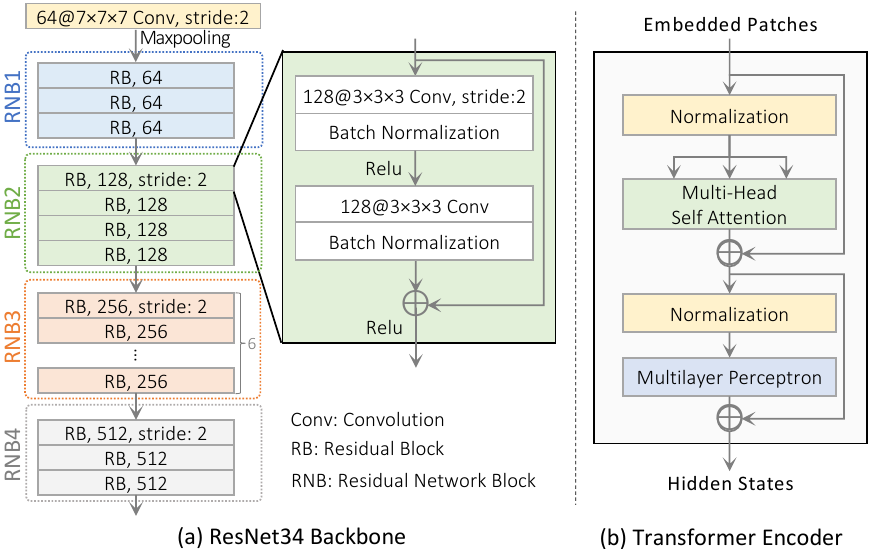}
\caption{The detailed architecture of (a) ResNet34 Backbone and (b) Transformer Encoder. The numbers 64, 128, 256, and 512 denote the channels of the corresponding convolution layers.}
\label{fig-RNB}
\end{figure}
\subsubsection{Data-Driven MRI Feature Learning}
To explore imaging biomarkers from MRIs in a data-driven manner, we propose to employ a ResNet34~\citep{he2016deep} as backbone for MRI feature learning. 
As shown in the top left panel of Fig.~\ref{fig-network}, the ResNet34 takes 3D MRI scans as input for feature learning. 
The detailed architecture of ResNet34 used in our work is shown in Fig.~\ref{fig-RNB}~(a).
The ResNet34 backbone contains a convolutional layer (kernel size: 7×7×7, stride: 2) and a max pooling layer (kernel size: 3×3×3, stride: 2),  followed by four residual network blocks (RNB). 
Each RNB consists of a stack of identical residual blocks. 
There are 3, 4, 6, and 3 residual blocks in the four RNBs, respectively. 
Each residual block has two convolutional layers, and each convolutional layer is followed by a batch normalization layer and a rectified linear unit (ReLU) activation function. 
A skip connection crosses these two convolution layers and adds the input directly before the final ReLU activation function. 
The length, width, and height dimensions are down-sampled with stride 2 between two adjacent RNBs. 
{\color{black}The channel numbers of convolution layers within each RNB are the same and decide the number of feature maps.} 
So, 512 feature maps are finally yielded by the ResNet34 backbone. 

{\color{black}
As shown in the upper middle part of Fig.~\ref{fig-network}, these feature maps are reshaped into small image patches with a resolution of ($l, w, h$) pixels through the following operations: 
\begin{itemize}
\vspace{-2mm}
\item Each feature map is reshaped into a vector ($x_p^n$) of patches with the dimension of $l\times w\times h$, where $n = 1, \cdots, N$ and $N$ denotes the number of patches. 
The size of the feature maps is small enough (i.e., $3\times 4\times 3$) due to the down-sampling between RNBs.
So, the final 512 feature maps are regarded as patches directly, and the $N=512$ here. 
\vspace{-2mm}
\item A sequence of embedded patches is generated by mapping the vector $x_p^n$ to $D$ dimensions through a trainable linear projection (i.e., 3D convolution). 
Since the input (i.e., embedded patches) and output (named hidden states) of the Transformer encoder are vectors of the same dimension, the dimension $D$ is also known as hidden size.
\vspace{-2mm}
\item A classification token $x_{class}$ is prepended to the sequence of embedded patches. 
The final hidden state of the classification token usually represents the aggregate representation for classification. 
When applied to our work, a better way of averaging or pooling the sequence of hidden states for the whole input embeddings is used ~\citep{devlin2018bert}.
\vspace{-2mm}
\item The embedded patches are further augmented with a one-dimensional position embedding $E_{pos}$, hence introducing positional information into the input, which is also learned during training. 
The embedded patches with position embedding can be represented as: $Z_0 = [x_{class}; x_p^1; \cdots; x_p^N] + E_{pos}$, where $E_{pos} = [0, \cdots, N]$ is the position embedding. 
\if false
\begin{equation}
\vspace{-8pt}
\centering
$Z = [x_{class}; x_p^1; \cdots; x_p^N] + E_{pos}$,
\vspace{-8pt}
\end{equation}
where $E_{pos} = [0; \cdots; N]$. 
\fi
\end{itemize}
}

Finally, the embedded patches $Z_0$ are fed into the subsequent feature fusion module. 
Note that the proposed HRL framework is very flexible, while other deep learning models can also be used as a backbone for feature extraction. 

\subsubsection{Handcrafted MRI Feature Extraction}
\label{S3_2_2}
The handcrafted MRI features are introduced for promoting the training of HRL and improving the performance of classification.
While features extracted by the ResNet backbone are generally based on MR image intensity, handcrafted MRI features could provide more structural information on local anatomical structures. 
Specifically, three types of handcrafted features are extracted from each MRI, including (1) structural statistics, such as size of surface area and gray matter volume, (2) attribute information like cortical thickness and mean curvature of the cortical surface, and (3) white matter parcellation information. 
These features can be calculated through MRI preprocessing with tissue segmentation and parcellation tools (i.e., FreeSurfer in our work), with details reported in Tables SI-SIX in \emph{Supplementary Materials}. 
The handcrafted features of each subject are concatenated as a 1,007-dimensional vector. 

As shown in the lower middle part of Fig. \ref{fig-network}, the handcrafted feature vector is reshaped to a list of vectors with a uniform length (i.e., the hidden size that is the same length as the embedded patches $x_p^n$). 
After reshaping, the list of vectors can be regarded as embedded patches $Z_1$ and fed into the Transformer encoder with $Z_0$ for feature fusion and abstraction.
Note that the data-driven MRI features need position embedding while the handcrafted features don't, since the 1D handcrafted features vector already eliminates position information.

\vspace{-2pt}
\subsubsection{Feature Fusion and Abstraction}
To effectively fuse the handcrafted features and data-driven MRI features, we propose using a Transformer encoder for hybrid feature fusion and abstraction. 
The Transformer encoder receives 1-dimensional data as input, so the embedded patches $Z_0$ and $Z_1$ are directly fed to the encoder. 
Due to the structure of the Transformer encoder, the output hidden states correspond to the input embedded patches one by one, and the size is the same (as shown in the right part of Fig. \ref{fig-network}). 

Fig. \ref{fig-RNB}(b) shows the detailed architecture of the Transformer encoder with three components.
(1) Multi-head self-attention (MSA) layer: This layer concatenates all the self-attention outputs linearly to embed information globally across the overall features. The attention heads help train local and global dependencies between features. 
(2) Multi-layer perceptron (MLP) Layer: The MLP contains a two-layer with Gaussian error linear unit. 
(3) Normalization Layer: This is added prior to MSA and MLP for improving the training time and overall performance. 
Moreover, residual connections are included after MSA and MLP. 
As a result, the Transformer encoder outputs a $D$-dimensional vector representation (i.e., hidden states in Fig.~\ref{fig-network}) for each embedded patch of the input sequence.

The Transformer encoder in HRL is used to fuse data-driven and handcrafted MRI features. 
Through its unique multi-head self-attention mechanism, long-term dependencies between different features can be established to realize feature fusion and representation learning ~\citep{zeng2022nlfftnet,sun2022transformer}. 
The encoder in Fig.~\ref{fig-RNB}~(b) usually are stacked on top of each other for feature extraction in computer vision tasks (typically stacking 6 or 12 of these blocks). 
But the Transformer encoder in HRL is used for feature fusion rather than feature extraction, only 1 iteration of the encoder is sufficient in our HRL. 
The hidden size and number of self-attention heads of the encoder are empirically set to $128$ and $16$, respectively, in the experiments.

\vspace{-2pt}
\subsubsection{Classification}
In order to quickly converge the model, we modified the original MLP head in ViT that is used for classification. 
The newly designed classification module consists of an average pooling layer and a linear layer with \emph{tanh} activation as non-linearity. 
The final hidden states from the Transformer encoder are averaged and fed to the linear layer for class prediction. 
The final outputs are converted to the probabilities that the subject belongs to different categories using the SoftMax function. 
Thus, a cross-entropy loss function is used for training the HRL. 

\subsection{Implementation}
The HRL is trained via a two-stage optimization strategy. 
(1) In the 1st stage, the ResNet can be trained, with 3D MRIs as input and their category labels as supervision. 
Then, we use the parameters of the trained model for the backbone initialization. 
(2) In the 2nd stage, we train the Transformer encoder (with ResNet backbone frozen), where handcrafted features and data-driven MRI features learned from ResNet are used as input and the category labels are treated as output. 
To balance the data, sMRI scans are duplicated and augmented using random affine transformation with TorchIO during the training stage. 
The Adam optimizer with a learning rate of ${10}^{-4}$ is used. 
{\color{black}We train the HRL with a maximum of 300 epochs, and an early stop strategy is applied when the prediction accuracy on training set exceeds the threshold (e.g., 0.9 in our work). 
The batch size is set as 4 with full-size MRIs as input due to the limitation of GPU memory.}
The experiments are implemented on PyTorch under the environment of Python 3.7 on the Ubuntu 18.04 system with a GPU (NVIDIA TITAN Xp) of 12 GB memory.

\section{Experiments}
\label{S4}
\subsection{Experimental Setup} 
\paragraph{\textbf{Evaluation Metrics}} 
Several metrics are used for performance evaluating, including the area under the ROC curve (AUC), accuracy (ACC), sensitivity (SEN), specificity (SPE), and F1-Score (F1s). 
A 5-fold cross-validation strategy is used in the experiments. 
Specifically, we first randomly split all subjects into 5 groups for each category. 
Then, one group is alternatively used as a test set and the remaining 4 groups are used as training set. 
The experiments are repeated 5 times independently to avoid any bias introduced by the random split. 
Two tasks are performed in the experiments, including (1) CN-D vs. CN-N classification, and (2) CI vs. CN-D vs. CN-N classification. 
In the binary task (\ie, CN-D vs. CN-N classification), both groups stay stable during the follow-up 5 years, and the mean and standard deviation of the evaluation metrics achieved in 5-fold cross validation are calculated. 
In the three-category task (\ie, CI vs. CN-D vs. CN-N classification), we use a well-known one-versus-all classification strategy~\citep{aly2005survey} and record all the prediction results of 5 folds and calculate the overall ACC and SEN results for each category. 
The confusion matrices are presented for detailed comparison and analysis.

\paragraph{\textbf{Competing Methods}} 
We compare the proposed HRL with the most popular machine learning methods used for depression detection and classification in LLD research, including support vector machine (\textbf{SVM})~\citep{kim2018application, mousavian2019depression, kambeitz2017detecting, saidi2020hybrid}, random forest (\textbf{RF})~\citep{lebedeva2017mri} and \textbf{XGBoost} (XGB)~\citep{chen2016xgboost, arun2018exploratory, arun2018boosted, sharma2020improving}). 
(1) In the SVM method, an SVM with Radial basis function kernel and regularization parameter $C=1.0$ is used for classification. 
(2) In the RF, the random forest classifier with 100 decision trees is used for classification. 
(3) In XGB, a grid search strategy is used to find a good combination of hyperparameters (i.e., number of boosting rounds, maximum tree depth for base learners, learning rate).
We use XGB with boosting rounds of 300, a maximum tree depth of 4, and a learning rate of 0.2.  
These three classifiers take handcrafted MRI features as inputs, including (1) average image intensities of gray matter (GM), white matter (WM), and cerebrospinal fluid (CSF) within pre-defined ROIs in AAL3 (denoted as {SVM/RF/XGB-GM, SVM/RF/XGB-WM} and {SVM/RF/XGB-CSF}, respectively) 
and (2) anatomical structural features extracted via FreeSurfer (denoted as {SVM/RF/XGB-FF}). 
For three-category problems, the most commonly used One-vs-the-rest (OvR) classification strategy is {\color{black}used~\citep{wu2006one}.}

We also compare the HRL with the state-of-the-art deep learning methods for MRI-based LLD research, including (1)  \textbf{ResNet}~\citep{he2016deep, yang2021deep}, (2) \textbf{Med3D}~\citep{chen2019med3d}, and (3) \textbf{ViT}~\citep{dosovitskiy2020image}. 

(1) {ResNet}: ResNet is a CNN-based architecture of a kind that stacks residual blocks on top of each other to form a network. The ResNet takes 3D MRI data as input, extracts feature through the stacked residual blocks, and makes the final class prediction using an average pooling layer and a fully connected layer. 
ResNet has many variants with similar architecture but different numbers of layers, the variants include ResNet18, ResNet34, ResNet50, ResNet101, ResNet152, and ResNet200. 
We choose the 3D version of ResNet18, ResNet34, and ResNet50 for comparison, considering our input size and GPU memory capacity. 
The Adam optimizer with a learning rate of ${10}^{-4}$, and a batch size of 4 is used for model training.
The trained ResNet34 model can be used for the ResNet34 backbone initialization in our HRL (i.e., the 1st stage optimization).

(2) {Med3D}: Med3D is a heterogeneous 3D network to co-train multi-domain 3D medical image segmentation so as to make a series of pretrained models. The Med3D pretrained models on large datasets can be transferred to new models in image segmentation and classification tasks and help improve the models' performance. We download the Med3D pretrained models through links in GitHub \footnote{https://github.com/Tencent/MedicalNet}, and transferred them to ResNet18, ResNet34, and ResNet50 for fine-tuning and performance comparison (denoted as {Med3D18/Med3D34/Med3D50}). The parameters setting for training Med3D is the same as ResNet.

(3) {ViT}: ViT consists of a stack of Transformer encoders for feature extraction and an MLP head for classification. 
The architecture of the Transformer encoder is the same as in HRL. 
For the convenience of comparison, we chose the 3D implementation of ViT (denoted as {ViT3D}). When the ViT model takes 3D MRI data as input, a large amount of GPU memory is required during training. 
For ViT3D, the inputs are down-sampled by $2\times2\times2$, and the iteration of the Transformer encoder is set to 6 due to the limits of GPU memory size (typically recommended set to 6 or 12 in 2D image classification in \citep{dosovitskiy2020image}). 
Other parameters as the hidden size and number of attention heads are set to 128 and 16 respectively, which is the same as in HRL.
The AdamW optimizer with a learning rate of ${10}^{-4}$, and average pooling in MLP is used for ViT training. The patch size is $8\times8\times8$, the batch size is 6, and the dimension of MLP is 512.

All competing deep learning models are trained with the sMRI data only and the same parameters setting of data preprocessing (e.g., histogram standardization, zero-mean normalization, intensity rescaling) and data augmentation as for HRL.

\begin{table}[!t]
\renewcommand{\arraystretch}{1}
\scriptsize
\centering
\caption{Results achieved by different methods (in terms of mean $\pm$ standard deviation) in the binary classification for LLD detection (i.e., CN-D vs. CN-N classification), with models trained on handcrafted features and MRI data. Methods marked as ``-FF'' denote that handcrafted Features extracted by FreeSurfer.}
\setlength\tabcolsep{3pt}
\begin{tabular}{lcccccccccc}
\toprule
~Method &AUC (\%) &ACC (\%) &SEN (\%) &SPE (\%) &F1s (\%)  \\ 
\hline
~SVM-GM  &58.14±4.94 &57.58±7.23 &56.34±14.76 &59.93±12.51 &63.11±9.21\\
~SVM-WM  &57.00±7.63 &56.42±6.42 &55.36±6.07 &58.63±12.87 &62.77±5.89\\
~SVM-CSF &54.63±6.30 &51.79±6.14 &46.12±6.04 &\textbf{63.14±6.99} &55.91±6.31\\
~SVM-FF  &53.43±3.66 &55.64±7.37 &60.38±20.46 &46.47±21.43 &63.13±10.01\\
~RF-GM  &51.70±4.50 &53.42±3.41 &56.48±8.92 &46.93±14.77 &61.43±5.16\\
~RF-WM  &52.84±7.46 &54.13±7.41 &56.46±9.94 &49.22±12.02 &61.78±8.29\\
~RF-CSF &50.34±4.98 &51.55±2.86 &53.75±2.62 &46.93±11.88 &59.62±1.28\\
~RF-FF  &\textbf{61.01±8.48} &\textbf{61.01±7.41} &60.40±14.62 &61.63±22.24 &\textbf{66.60±9.13}\\
~XGB-GM  &55.27±4.45 &59.57±5.73 &68.13±8.71 &42.42±5.37 &68.95±5.61\\
~XGB-WM  &51.25±5.16 &56.88±6.35 &68.06±9.51 &34.44±6.82 &67.50±6.67\\
~XGB-CSF &46.67±5.31 &50.75±4.82 &58.96±5.44 &34.38±7.57 &61.38±4.09\\
~XGB-FF  &52.55±6.71 &56.83±8.46 &\textbf{64.97±15.93} &40.13±15.13 &65.89±10.34\\
\hline
~ResNet18 &59.44$\pm$7.69 &62.34$\pm$5.39 &67.77$\pm$8.69 &51.11$\pm$19.29 &70.61$\pm$4.54 \\
~ResNet34 &59.44$\pm$6.45 &62.34$\pm$6.37 &67.77$\pm$8.91 &51.11$\pm$11.73 &70.54$\pm$6.10 \\
~ResNet50 &55.31$\pm$4.28 &57.49$\pm$5.76 &61.66$\pm$12.93 &48.95$\pm$13.78 &65.44$\pm$8.75 \\
~Med3D18 &57.12$\pm$6.97 &61.20$\pm$4.88 &{68.88$\pm$3.62} &45.35$\pm$14.31 &{70.48$\pm$3.17} \\
~Med3D34 &{61.91$\pm$5.22} &{64.98$\pm$8.31} &{70.55$\pm$14.24} &\textbf{53.26$\pm$6.10} &{72.44$\pm$8.50} \\
~Med3D50 &58.57$\pm$7.09 &62.71$\pm$8.03 &{70.55$\pm$9.93} &46.60$\pm$6.01 &71.52$\pm$7.41 \\
~ViT3D &54.28±11.13&58.94±11.66&67.78±13.97&40.78±12.81&68.48±10.42\\ 
~HRL~(Ours) &\textbf{62.75±7.23} & \textbf{66.45±5.18} &\textbf{73.33±10.32} &{52.16±20.19} &\textbf{74.36±4.77}\\
\bottomrule
\end{tabular}

\label{MDD2_comparison}
\end{table}

\subsection{Results of Binary Classification  }
The results achieved by the proposed HRL and the competing methods in binary classification task (\ie, CN-D vs. CN-N) are reported in Table~\ref{MDD2_comparison}. 
From Table~\ref{MDD2_comparison}, we have the following observations. 
\emph{First}, our HRL method outperforms conventional machine learning methods (i.e., SVM, RF, and XGB) and three deep learning methods (i.e., ResNet, Med3D and ViT3D) in most cases. 
Especially, the HRL achieves the highest SEN value (i.e., $73.33\%$) which may be very useful to accurately identify subjects with depression. 
These results suggest the effectiveness of the proposed hybrid learning framework for LLD identification. 
\emph{Second}, deep learning methods usually perform a little better than machine learning methods. 
This implies that incorporating MRI feature learning and classification into a unified framework via deep learning helps boost the identification performance, compared to the twelve machine learning methods that treat MRI feature extraction and classification as standalone tasks. 
\emph{Besides}, among the seven competing deep learning methods, Med3D34 performs overall better than others. 
The possible reason is that using complex network architecture (e.g., ViT3D) or using deeper layers (i.e., ResNet50 and Med3D50) may not necessarily boost the learning performance due to the limited number of training samples. 
\emph{In addition}, machine learning methods (i.e., SVM-FF, RF-FF and XGB-FF) with anatomical features usually yield higher SEN values compared to their counterparts (e.g., SVM-CSF, RF-CSF and XGB-CSF) with image intensity-based features. 
This may due to the fact that features extracted via FreeSurfer contain more information about brain structures, which helps to identify anatomical differences between CN-D and CN-N in brain MRI.

\begin{table}[!t]
\renewcommand{\arraystretch}{1}
\scriptsize
\centering
\caption{Accuracy (ACC) results achieved by different methods in three-category classification (i.e., CI vs. CN-D vs. CN-N), with models trained on handcrafted features and MRI data. Methods marked as "-FF" denote that handcrafted Features extracted by FreeSurfer.}
\setlength\tabcolsep{7pt}
\begin{tabular}{lc cc c }
\toprule
~Method &ACC (\%) &ACC$_{CN-N}$ (\%) &ACC$_{CN-D}$ (\%) &ACC$_{CI}$ (\%)\\ 
\hline
~SVM-GM  &40.14 &51.70 &47.62 &80.95\\
~SVM-WM  &42.86 &53.40 &50.68 &81.63\\
~SVM-CSF &39.12 &50.68 &47.28 &80.27\\
~SVM-FF  &37.41 &48.98 &47.28 &78.57\\
~RF-GM   &53.40 &64.63 &56.46 &85.71\\
~RF-WM   &53.40 &62.24 &55.44 &\textbf{89.12}\\
~RF-CSF  &53.06 &63.27 &56.12 &86.73\\
~RF-FF   &\textbf{55.44} &\textbf{65.99} &\textbf{57.82} &87.07\\
~XGB-GM  &50.68 &63.61 &53.74 &84.01\\
~XGB-WM  &52.38 &62.59 &54.42 &87.76\\
~XGB-CSF &46.26 &57.82 &49.66 &85.03\\
~XGB-FF  &54.42 &64.97 &56.80 &87.07\\
\hline
~ResNet18 &55.78 &66.67 &61.56 &83.33\\
~ResNet34 &51.36 &65.31 &57.82 &79.59\\
~ResNet50 &54.08 &62.24 &\textbf{61.90} &84.01\\
~Med3D18 &50.00 &60.54 &56.12 &83.33\\
~Med3D34 &52.04 &66.67 &60.54 &76.87\\
~Med3D50 &51.02 &61.56 &57.82 &82.65\\
~ViT3D    &49.32 &62.93 &58.16 &77.55\\
~HRL~(Ours) &\textbf{57.48} &\textbf{69.05} &61.56 &\textbf{84.35}\\
\bottomrule
\end{tabular}
\label{LLD3_ACC}
\end{table}
\subsection{Results of Three-Category Classification }
\begin{figure*}[!t]
\setlength{\belowdisplayskip}{-1pt}
\setlength{\abovedisplayskip}{-1pt}
\setlength{\abovecaptionskip}{-1pt}
\setlength{\belowcaptionskip}{-1pt}
\centering
\includegraphics[width=0.99\textwidth]{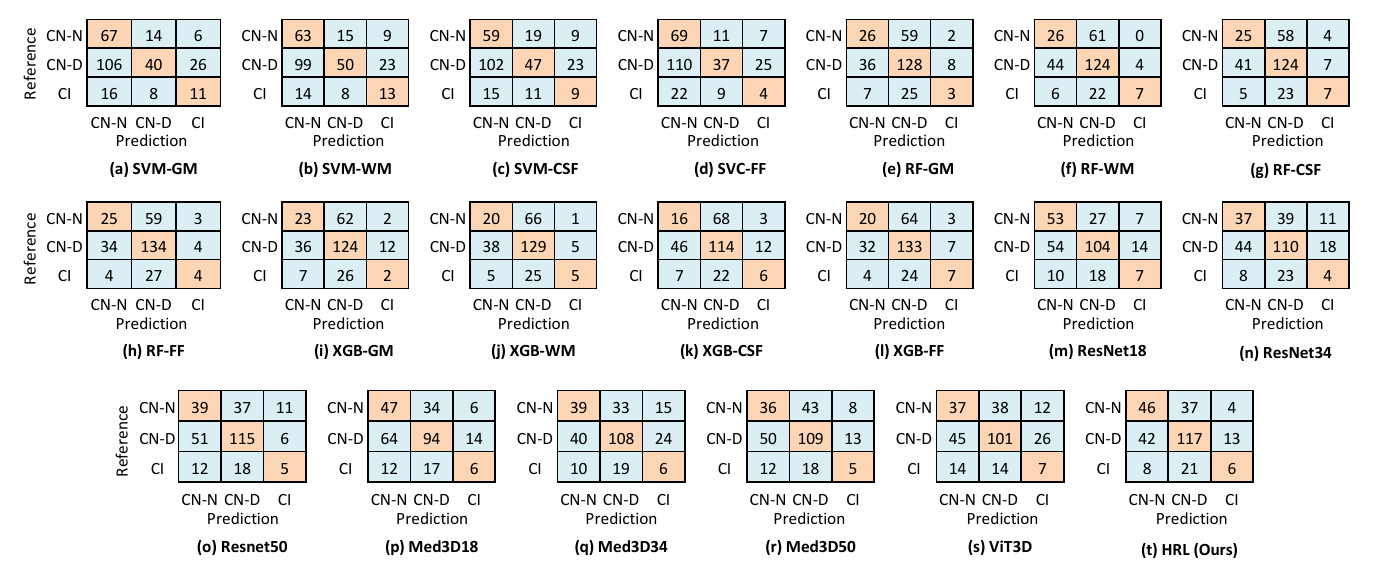}
\caption{Confusion matrices achieved by different methods in three-category classification (i.e., CI vs. CN-D vs. CN-N).} 
\label{LLD3-confusion-matrix}
\end{figure*}

We further perform diagnostic outcome prediction in LLD via a three-category classification task, \ie, CI vs. CN-D vs. CN-N classification. 
The ACC and SEN values of each category achieved by different methods in this task are reported in Table~\ref{LLD3_ACC} and Table~\ref{LLD3_SEN}, respectively. 
The confusion matrices achieved by different methods are shown in Fig.~\ref{LLD3-confusion-matrix}. 

Table~\ref{LLD3_ACC} suggests that our HRL achieves the best overall ACC result, while RF-FF yields the second-best ACC value. 
Table~\ref{LLD3_SEN} suggests that the HRL generates the highest SEN value for the CN-D category among the deep learning methods. 
Besides, even our method can achieve good ACC values for the CI category when compared to the other two classes (see Table~\ref{LLD3_ACC}), but the sensitivity for CI is not good (see Table~\ref{LLD3_ACC}). 
Similarly, SVM-based models generally obtain high SEN scores for CN-N but not good SEN results for CN-D and CI. 
A reasonable explanation is that the number of CI subjects is very small (\ie, 25) and even fewer for the test in 5-fold cross validation. 
On the other hand, from Tables~\ref{LLD3_ACC}-\ref{LLD3_SEN} and Table~\ref{MDD2_comparison}, we can see that the overall performance of each model drops dramatically as the number of categories increases.
The low recognition rate of the newly added category (\ie, CI) greatly reduces the overall performance of each predictive model.

The confusion matrices in Fig.~\ref{LLD3-confusion-matrix} show more detailed and specific performance of each method on the three different categories.
Fig.~\ref{LLD3-confusion-matrix} suggests that there is a data imbalance issue in the three-category classification task, while deep learning methods generally outperform conventional machine learning methods in terms of the overall accuracy. 
The most possible reason is that machine learning methods use handcrafted MRI features whose dimensionality and discriminative power are limited by templates and segmentation tools, while deep learning models can extract MRI features tailored for downstream tasks and thus achieve better classification performance due to the homogeneity between MRI features and the prediction model. 
Also, our HRL performs reasonably better than the other deep learning methods in general. 
The above experimental results and analysis imply that the main limiting factor for the application of deep learning in diagnostic outcome prediction of LLD could be the problem of data imbalance between different categories.

\begin{table}[!t]
\renewcommand{\arraystretch}{1}
\scriptsize
\centering
\caption{Sensitivity (SEN) results achieved by different methods in three-category classification (i.e., CI vs. CN-D vs. CN-N), with models trained on handcrafted features and MRI data. Methods marked as "-FF" denote that handcrafted Features extracted by FreeSurfer.}
\setlength\tabcolsep{7pt}
\begin{tabular}{lc cc c }
\toprule
~Method &SEN (\%) &SEN$_{CN-N}$ (\%) &SEN$_{CN-D}$ (\%) &SEN$_{CI}$ (\%)\\ 
\hline
~SVM-GM  &43.90 &\textbf{77.01} &23.26 &31.43\\
~SVM-WM  &\textbf{46.21} &72.41 &29.07 &\textbf{37.14}\\
~SVM-CSF &40.29 &67.82 &27.33 &25.71\\  
~SVM-FF  &37.42 &79.31 &21.51 &11.43\\
~RF-GM   &37.63 &29.89 &74.42 & 8.57\\
~RF-WM   &40.66 &29.89 &72.09 &20.00\\
~RF-CSF  &40.28 &28.74 &72.09 &20.00\\
~RF-FF   &39.36 &28.74 &\textbf{77.91} &11.43\\
~XGB-GM  &34.75 &26.44 &72.09 & 5.71\\
~XGB-WM  &37.42 &22.99 &75.00 &14.29\\
~XGB-CSF &33.94 &18.39 &66.28 &17.14\\
~XGB-FF  &40.10 &22.99 &77.33 &20.00\\
\hline
~ResNet18 &\textbf{47.13} &\textbf{60.92} &60.47 &\textbf{20.00}\\
~ResNet34 &39.30 &42.53 &63.95 &11.43\\
~ResNet50 &41.99 &44.83 &66.86 &14.29\\
~Med3D18 &41.94 &54.02 &54.65 &17.14\\
~Med3D34 &41.59 &44.83 &62.79 &17.14\\
~Med3D50 &39.68 &41.38 &63.37 &14.29\\
~ViT3D    &40.42 &42.53 &58.72 &\textbf{20.00}\\
~HRL~(Ours) &46.01 &52.87 &\textbf{68.02} &17.14\\
\bottomrule
\end{tabular}
\label{LLD3_SEN}
\end{table}

\begin{figure*}[!t]
\setlength{\belowdisplayskip}{-1pt}
\setlength{\abovedisplayskip}{-1pt}
\setlength{\abovecaptionskip}{-1pt}
\setlength{\belowcaptionskip}{-1pt}
\centering
\includegraphics[width=1\textwidth]{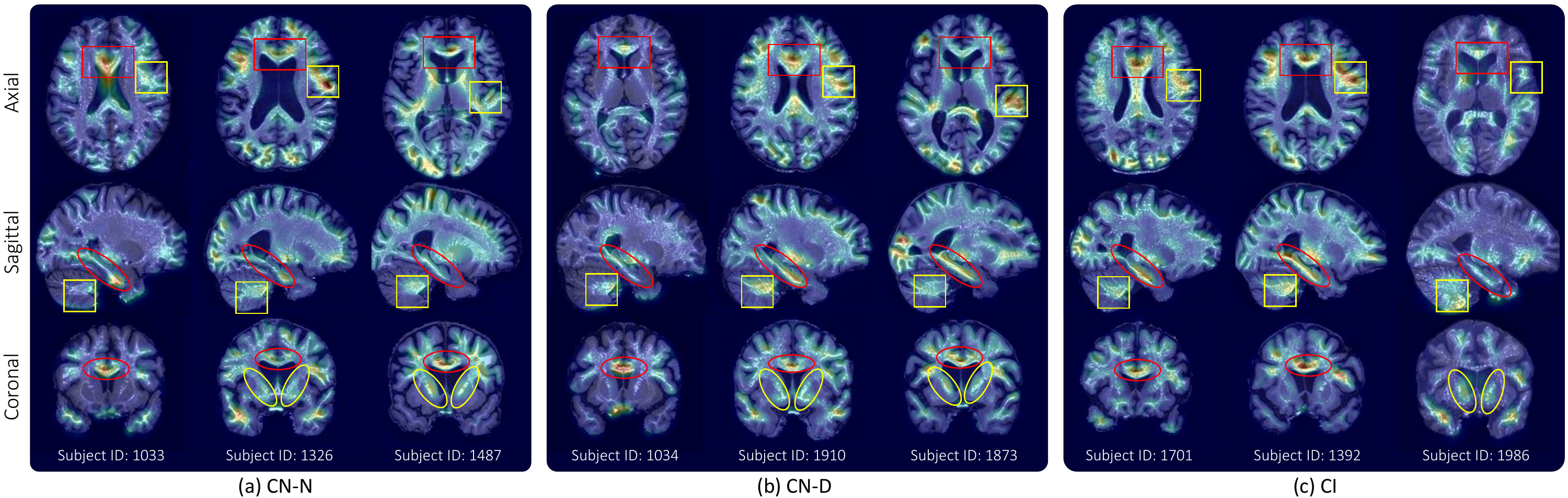}
\caption{Feature maps extracted by the ResNet backbone in our HRL model in three-category classification. The boxes and ellipses mark the most informative ROIs defined with the AAL3 template. (Red boxes in the axial view: ACC\_pre\_L and ACC\_pre\_R; Yellow boxes in the axial view: Frontal\_Inf\_Oper\_L; Red ellipses in the sagittal view: Hippocampus\_L or Hippocampus\_R; Yellow boxes in the sagittal view: Cerebellum; Red ellipses in the coronal view: ACC\_sup\_L and ACC\_sup\_R; Yellow ellipses in the coronal view: Caudate\_L and Caudate\_R).}
\label{fig_featuremap}
\end{figure*}

\subsection{Learned Feature Maps}
The feature maps extracted by the first residual network block of ResNet in our HRL in the task of three-category classification are shown in Fig.~\ref{fig_featuremap}, with the most discriminative ROIs marked by boxes and ellipses. 
For each category, we show three representative subjects and the learned feature maps in the axial, sagittal, and coronal views. 

As shown in Fig.~\ref{fig_featuremap}, the most discriminative ROIs (marked by boxes and ellipses) include ACC\_pre\_L and ACC\_pre\_R (i.e., anterior cingulate cortex, subgenual and pregenual, red boxes in the axial view), Frontal\_Inf\_Oper\_L (i.e., the opercular part of left inferior frontal gyrus, yellow boxes in the axial view), Hippocampus\_L and Hippocampus\_R (i.e., hippocampus, red ellipses in the sagittal view), Caudate\_L and Caudate\_R (i.e., caudate, yellow ellipses in the coronal view), ACC\_sup\_L and ACC\_sup\_R (i.e., anterior cingulate cortex, supracallosal, red ellipses in the coronal view), cerebellum (yellow boxes in the sagittal view). 
These findings are consistent with previous studies ~\citep{zhang2018brain, yang2021deep}, suggesting that our HRL has great potential to discover the most discriminative ROIs that contribute to diagnostic outcome prediction.

\begin{figure*}[!t]
\setlength{\belowdisplayskip}{0pt}
\setlength{\abovedisplayskip}{0pt}
\setlength{\abovecaptionskip}{0pt}
\setlength{\belowcaptionskip}{0pt}
\centering
\includegraphics[width=0.99\textwidth]{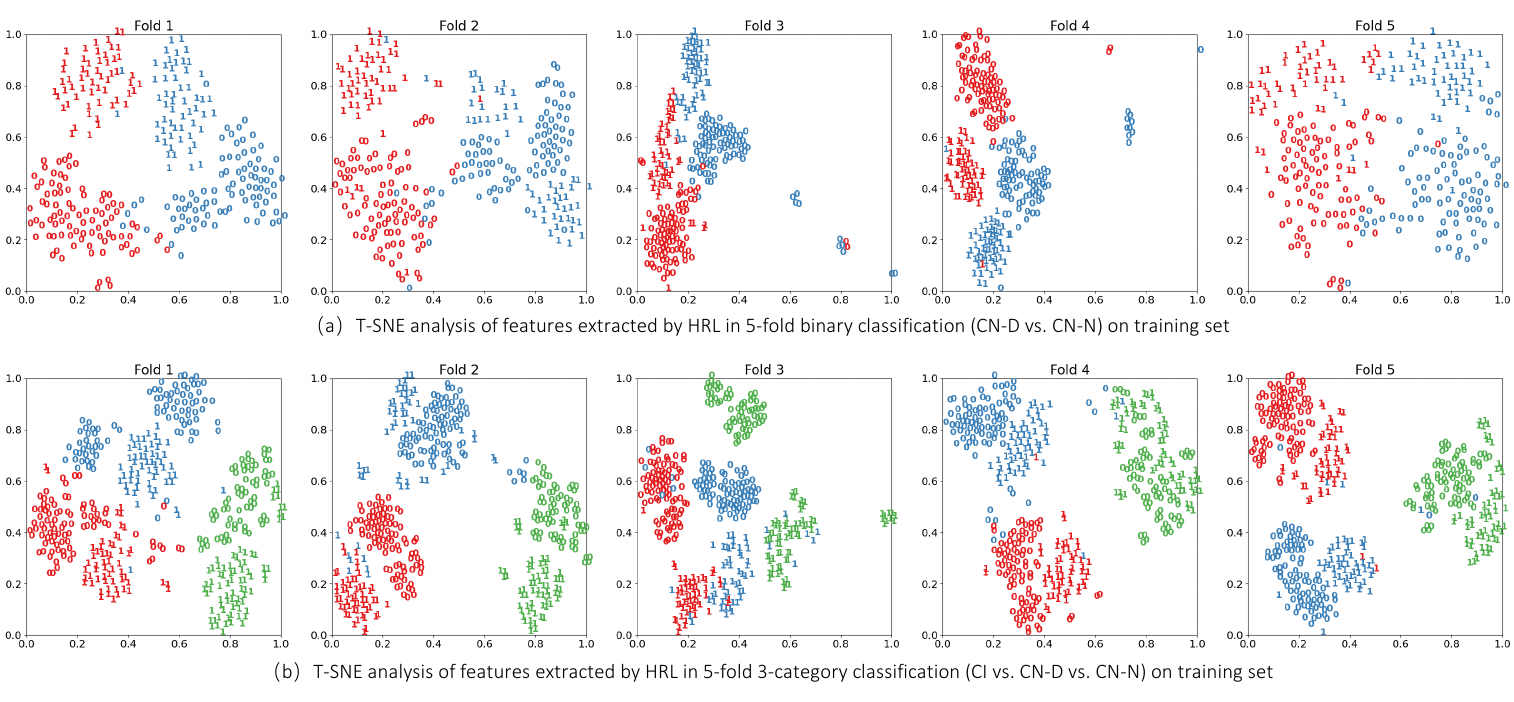}
\caption{Feature distributions of the ResNet backbone in HRL using t-SNE on the training set. Results of 5 folds in (a) binary classification and (b) three-category classification are shown. The red, blue, and green colors denote CN-N, CN-D, and CI categories, respectively, {\color{black}while 0 and 1 denote the imaging site of NCODE and NBOLD, respectively.}}
\label{fig_tSNE}
\end{figure*}

\subsection{Visualization of Data-Driven MRI Features}
To better understand results in Tables~\ref{MDD2_comparison}-\ref{LLD3_SEN} and the MRI features extracted by HRL, the t-SNE method~\citep{van2008visualizing} is used to analyze the distribution of intermediate MRI features in the binary and three-category classification tasks, with results reported in Fig.~\ref{fig_tSNE}. 
The features generated by the last layer of the ResNet backbone of HRL in 5 folds on training data are extracted for analysis. 
In Fig.~\ref{fig_tSNE}, different colors denote different categories (i.e., red for CN-N, blue for CN-D, green denotes for CI) and different digits correspond to different datasets (i.e., 0 for NCODE and 1 for NBOLD).

It can be observed from Fig.~\ref{fig_tSNE} that the learned MRI features are generally well separated by category labels and datasets on the training set. 
But the performance of our HRL on the test set are not that promising, as shown in Tables~\ref{MDD2_comparison}-~\ref{LLD3_SEN}. 
This may due to the significant inter-dataset differences in the NCODE and NBOLD studies, which is a common problem in multi-site studies that negatively affect generalization ability of deep learning models~\citep{guan2021domain}. 
This effect is pronounced for HRL with limited sample size. 
Designing advanced strategies to reduce such inter-dataset data distribution differences is expected to further promote the prediction performance.

\vspace{-8pt}
\section{Discussion}
\label{S5}

{\color{black}
\if false
\subsection{Comparison with Previous Studies}
In this work, we investigate the 5-year cognitive progression of LLD based on T1-weighted MRI and diagnosis information of long follow-up time. 
Experimental results in both tasks of  CN-D vs. CN-N and CI vs. CN-D vs. CN-N classification demonstrate the effectiveness of the proposed methods.  
Besides, from the confusion matrices in Fig.~\ref{LLD3-confusion-matrix}, it can be seen that the classification performance is imbalanced between categories with machine learning models. 
The most possible reason is that the dimension of the handcrafted features is limited by the atlas template and segmentation tools.  
The deep learning model can extract MRI features according to downstream task, and use the visualization of the feature map to help discover potential features and the association between features.
\fi

\subsection{Interpretation of Identified MRI Biomarkers}
In this work, we investigate the 5-year cognitive progression of LLD based on T1-weighted MRI and diagnosis information of long follow-up time. 
From the feature maps in Fig.~\ref{fig_featuremap}, we can see the most discriminative brain regions identified by our method include anterior cingulate cortex (ACC), frontal lobe, hippocampus, caudate, and cerebellum. 
These brain regions are generally consistent with those in previous studies related to depression~\citep{zhang2018brain,banasr2021macro,toenders2022association,yang2021deep}. 
For example, Zhang et al.~\citep{zhang2018brain} reviewed the previous findings on brain structural changes in depression and found significant alterations in several brain regions such as the frontal lobe, hippocampus, temporal lobe, thalamus, striatum, and amygdala in patients with major depressive disorder. 
Barnas et al.~\citep{banasr2021macro} focused on the study of prefrontal cortex and hippocampus because these brain regions have been found in many previous studies of depression to be associated with decreased volume of gray matter.
A recent study found that depression severity was associated with a thinner rostral ACC and melancholic depressive symptoms were negatively associated with caudal ACC thickness~\citep{toenders2022association}.
Yang et al.~\citep{yang2021deep} reported that the corpus callosum, hippocampus, cerebellum, insular, caudate nucleus, and brain steam are the most discriminative regions for depression symptom factors regression via deep learning. 
From the comparison, we can see that the identified discriminative brain regions in MRI could be used as potential MRI biomarkers for the assessment of cognitive change in LLD subjects with 5 years of a cognitively normal diagnosis, which is a time frame in which preventative or disease-modifying treatments may be most effective.
}

\subsection{Ablation Study}
\begin{table}[!t]
\renewcommand{\arraystretch}{1.1}
\scriptsize
\centering
\caption{Ablation study in binary classification task (i.e., CN-D vs. CN-N), with MRI down-sampled by 2×2×2 for training efficacy. The terms ``-H'' and ``-D'' denote models with handcrafted MRI features extracted by FreeSurfer and data-driven MRI features via the ResNet backbone, respectively.}
\setlength\tabcolsep{4pt}
\begin{tabular}{lc cc cc }
\toprule
~Method &AUC (\%) &ACC (\%) &SEN (\%) &SPE (\%) &F1s (\%)\\ 
\hline
~HRL-H     &52.75±5.54 &56.07±5.67 &62.83±15.55 &42.68±21.47 &64.77±8.20\\
~HRL-D  &56.16±4.13 &58.47±5.55 &\textbf{62.97±11.18} &49.35±9.82 &66.42±7.55\\
~HRL &\textbf{59.06±8.84} &\textbf{60.14±6.60} &61.97±14.61 &\textbf{56.14±25.91} &\textbf{66.72±8.82}\\
\bottomrule
\end{tabular}
\label{ablation}
\end{table}

The proposed HRL contains a Transformer encoder for fusion of data-driven and handcrafted MRI features.  
To investigate the influence of the transformer, we compare the HRL (with ResNet34 as the backbone) with its variants for ablation analysis, including \textbf{HRL-H} that uses handcrafted MRI features only and \textbf{HRL-D} with data-driven MRI features via ResNet only.
For a fair comparison, we set their key hyperparameters (e.g., learning rate and batch size) to be the same as HRL. 
All models are initialized with pretrained Med3D parameters and retrained without any limitations. 
Due to the limitation of GPU memory, we down-sampled the input MRI scans by 2×2×2 for training efficiency. 
The results of our HRL and its variants in the binary classification task are reported in Table~\ref{ablation}. 

It can be seen from Table~\ref{ablation} that our HRL achieves the overall best performance than its two degenerate variants (i.e., HRL-M and HRL-D). 
These results suggest that fusing handcrafted and data-driven MRI features via the Transformer encoder (as we do in HRL) is effective in identifying CN-D from CN-N subjects, compared to methods that use only handcrafted or data-driven MRI features. 
Besides, HRL-D outperforms HRL-M with a large margin. For instance, HRL-D yields an AUC of 56.16\%, which is 3.91\% higher than that of HRL-H. 
This demonstrates that data-driven MRI features learned by the ResNet backbone may be more discriminative in detecting control normal subjects with depression compared to handcrafted MRI features. 

\begin{figure}[!t]
\setlength{\belowdisplayskip}{0pt}
\setlength{\abovedisplayskip}{0pt}
\setlength{\abovecaptionskip}{0pt}
\setlength{\belowcaptionskip}{0pt}
\centering
\includegraphics[width=0.47\textwidth]{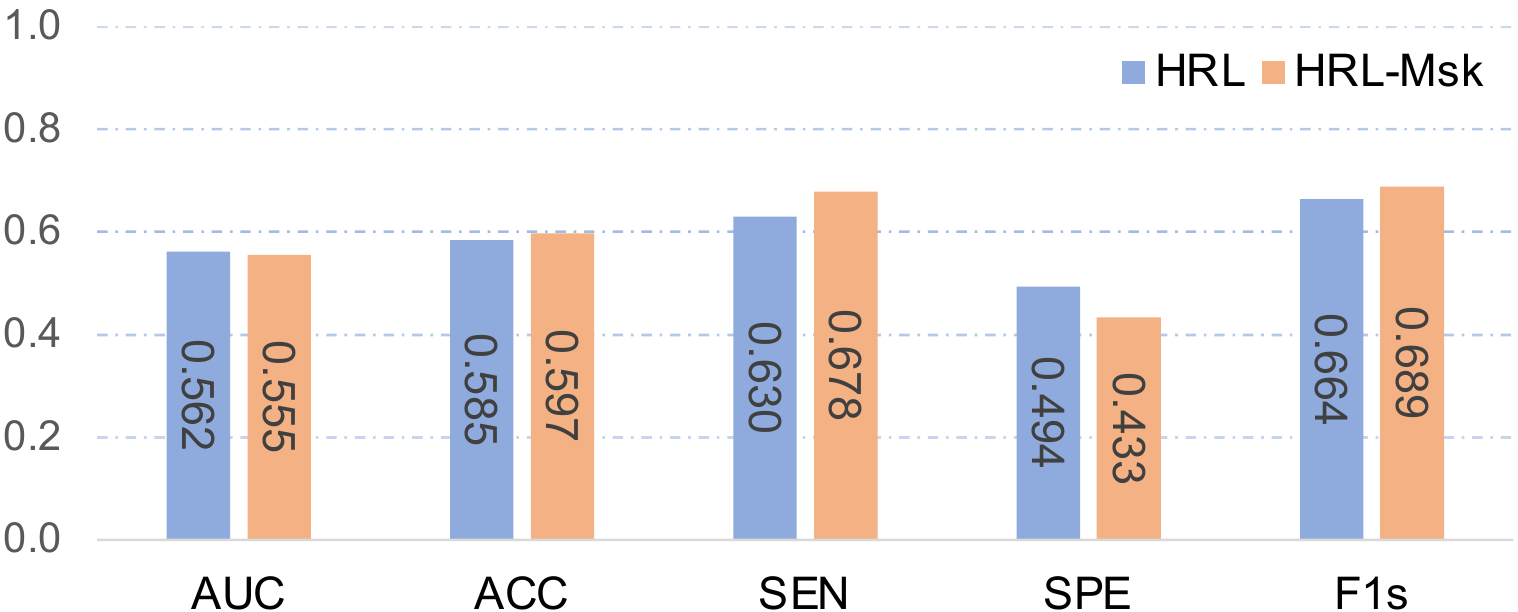}
\caption{Performance comparison of HRL and its variant HRL-Msk that use a predefined ROI mask in CN-D vs. CN-N classification. A total of 34 ROIs defined in AAL3 are selected by two experienced radiologists and clinicians to mask the input MRI scans.} 
\label{fig_ROI_MSK}
\end{figure}


\subsection{Influence of Predefined ROI Mask}
Based on the analysis results in Fig.~\ref{fig_featuremap}, we can see that some brain regions may be more informative for diagnostic outcome prediction. 
So, we perform an additional experiment to validate whether using predefined ROI masks help improve the prediction performance. 
A total of 34 potential LLD-associated ROIs (defined in the AAL3 template) are selected by two experienced radiologists and clinicians, including ACC (i.e., anterior cingulate cortex), OFC (i.e., orbitofrontal cortex), prefrontal cortex, hippocampus, caudate, insular, cingulate, amygdala, and nuclear acumens. 
The names and IDs of these ROIs are given in Table SX of the \emph{Supplementary Materials}.
These selected ROIs are used as a ROI mask of the input MRIs for model training and test. 
Results of our HRL and its variant  with the predefined ROI mask (denote as HRL-Msk) are reported in Fig.~\ref{fig_ROI_MSK}. 
The results in Fig.~\ref{fig_ROI_MSK} show that the performance of HRL-Msk is comparable with that of HRL, even when most of the brain regions are masked out in MRIs. 
It indicates that these predefined ROIs contain considerable information for the classification task.

\if false
The pre-selected 34 ROIs:
5 Frontal_Mid_2_L 5
6 Frontal_Mid_2_R 6
7 Frontal_Inf_Oper_L 7
8 Frontal_Inf_Oper_R 8
21 Frontal_Med_Orb_L 21
22 Frontal_Med_Orb_R 22
23 Rectus_L 23
24 Rectus_R 24
25 OFCmed_L 25
26 OFCmed_R 26
27 OFCant_L 27
28 OFCant_R 28
29 OFCpost_L 29
30 OFCpost_R 30
31 OFClat_L 31
32 OFClat_R 32
33 Insula_L 33
34 Insula_R 34
39 Cingulate_Post_L 39
40 Cingulate_Post_R 40
41 Hippocampus_L 41
42 Hippocampus_R 42
45 Amygdala_L 45
46 Amygdala_R 46
75 Caudate_L 75
76 Caudate_R 76
151 ACC_sub_L 151
152 ACC_sub_R 152
153 ACC_pre_L 153
154 ACC_pre_R 154
155 ACC_sup_L 155
156 ACC_sup_R 156
157 N_Acc_L 157
158 N_Acc_R 158
\fi

\begin{figure}[!t]
\setlength{\belowdisplayskip}{0pt}
\setlength{\abovedisplayskip}{0pt}
\setlength{\abovecaptionskip}{0pt}
\setlength{\belowcaptionskip}{0pt}
\centering
\includegraphics[width=0.47\textwidth]{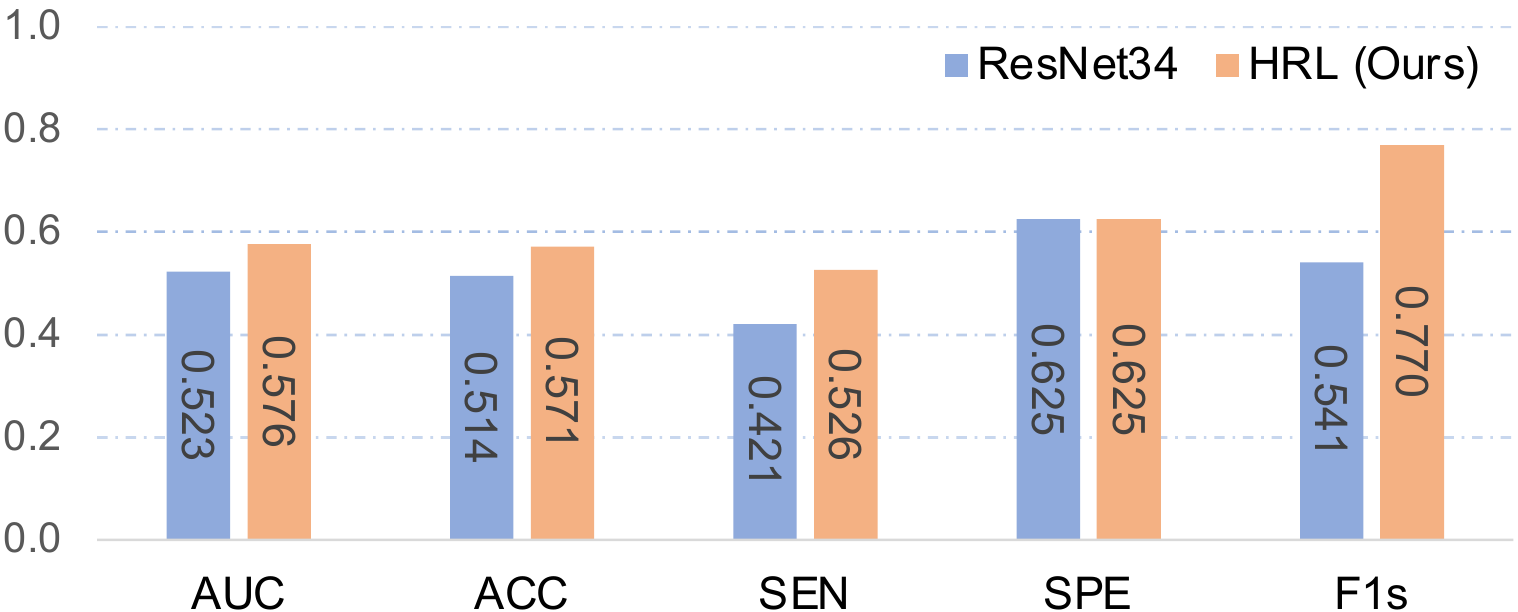}
\caption{Results of cross-dataset CI-to-AD diagnostic outcome prediction (i.e., CIND vs. AD classification). The models are trained for binary classification (i.e., AD vs. CN) with MRI selected from ADNI and transferred to CIND vs. AD classification.}
\label{fig-LLD2AD}
\end{figure}

\if false
\begin{table}[!t]
\setlength{\belowdisplayskip}{-1pt}
\setlength{\abovedisplayskip}{-1pt}
\setlength{\abovecaptionskip}{-1pt}
\setlength{\belowcaptionskip}{-1pt}
\setlength{\tabcolsep}{0.65mm}
\centering
\footnotesize
\caption{Results of Cross-Dataset CI-to-AD diagnostic outcome prediction (i.e., CIND vs. AD classification).}
\setlength\tabcolsep{3pt}
\begin{tabular}{lccccc}
\toprule
~Method   &AUC (\%) &ACC (\%) &SEN (\%) &SPE (\%) &F1s (\%)  \\ \hline
~{ResNet34} &52.30 &51.43 &42.11 &\textbf{62.50} &54.05\\
~HRL (Ours) &\textbf{57.57} &\textbf{57.14} &\textbf{52.63} &\textbf{62.50} &\textbf{57.14}\\ \bottomrule
\end{tabular}
\label{tab-LLD2AD}
\end{table}
\fi

\subsection{Cross-Dataset CIND-to-AD Prediction}
\label{S5.4}
Due to the limited number of CIND and AD subjects in this study, we cannot train a learning model for CIND-to-AD diagnostic outcome prediction based on the current data. 
To address this issue, we further propose a transfer learning strategy by training a model on the large-scale ADNI dataset~\citep{jack2008alzheimer} and testing it on subjects in this study. 
Inspired by ~\citep{cheng2015domain}, we use CN and AD subjects as auxiliary domain rather than MCI and AD subjects for the training of HRL.
Specifically, a total of 795 subjects (i.e., 436 CN and 359 AD subjects) from ADNI are used for model training. 

And 20 subjects of each category in ADNI are randomly selected for validation and the remaining are used for training the model to identify AD patients from cognitively normal subjects. 
Then, we perform AD vs. CIND classification on subjects in this study using the model trained on ADNI directly.  
In this experiment, we compare the proposed HRL with ResNet34, where HRL fuse handcrafted and data-driven MRI features and ResNet34 only use MRI as input.
The results of our HRL and ResNet34 in AD vs. CIND classification are reported in Fig.~\ref{fig-LLD2AD}.

It can be observed from Fig.~\ref{fig-LLD2AD} that both HRL and ResNet34 produce reasonable results in the challenging task, even though there exists a semantic gap between category labels in these two studies (i.e., CN \& AD in ADNI, and CIND \& AD in this work).
This suggests that using large-scale MRI data for model training could help to improve the performance of deep learning methods. 
In addition, our HRL generally outperforms ResNet34 in predicting the diagnostic outcome of CIND-to-AD, further validating the necessity of fusing data-driven and hand-crafted MRI features (as we do in HRL).

\begin{table}[!t]
\renewcommand{\arraystretch}{1.1}
\scriptsize
\centering
\caption{Results achieved by HRL with different training strategies in binary category classification (i.e., CN-D vs. CN-N), with MRI down-sampled by 2×2×2 for training efficacy. HRL-S denotes the model trained from scratch. HRL-R uses pretrained ResNet for network parameter initialization, and is then fine-tuned by joint training of ResNet and Transformer encoder. 
}
\setlength\tabcolsep{3pt}
\begin{tabular}{lc cc cc }
\toprule
~Method &AUC (\%) &ACC (\%) &SEN (\%) &SPE (\%) &F1s (\%)\\ 
\hline
~HRL-S  &60.11±10.51 &58.52±9.25 &54.93±10.91 &\textbf{65.29±18.39} &63.47±9.64\\
~HRL-R  &59.06±8.84 &60.14±6.60 &61.97±14.61 &56.14±25.91 &66.72±8.82\\
~HRL  &\textbf{61.12±8.05} &\textbf{65.77±7.95} &\textbf{75.11±8.02} &47.12±9.31 &\textbf{74.37±6.69}\\
\bottomrule
\end{tabular}
\label{strategy}
\end{table}

\subsection{Influence of Two-Stage Optimization Strategy}
To validate the effectiveness of the proposed two-state optimization strategy, we further compare the HRL with its two variants (called \textbf{HRL-S} and \textbf{HRL-R}) that use different training solutions. 
Specifically, the HRL-S is trained from scratch, without pretraining the ResNet backbone. 
Similar to our HRL, the HRL-R also uses a two-stage optimization strategy. 
The difference is that, in the 2nd stage, HRL-R jointly trains the ResNet backbone and the Transformer encoder for classification, while our HRL only train the Transformer with the ResNet backbone frozen. 
The comparison results are shown in Table~\ref{strategy}. Table~\ref{strategy} suggests that the proposed two-stage training strategy helps promote the performance of HRL. 
Also, the joint training of ResNet and Transformer in HRL-R cannot produce improved results, possibly due to the heterogeneity between the ResNet and Transformer architectures.

\subsection{Limitations and Future Work}
Several issues need to be considered in future studies. 
\emph{First}, the impact of data imbalance is more evident under the limited sample size. 
We will investigate more suitable feature extraction models or training strategies (e.g., meta-learning~\citep{vilalta2002perspective}, unsupervised constructive learning~\citep{pankajavalli2022independent}) to address the issues of small-sample-size and imbalanced data.
\emph{Besides}, there exists significance inter-site difference between NCODE and NBOD studies, as can be observed through t-SNE analysis in Fig.~\ref{fig_tSNE}. 
We will design advanced data harmonization/adaptation methods~\citep{kamnitsas2017unsupervised,guan2021domain} to reduce the data heterogeneity between different sites. 
\emph{Furthermore}, we use only structural MRI in this work, 
without considering other modalities such as functional MRI~\citep{pilmeyer2022functional} and demographic information~\citep{nayak2022socio}. 
In the future, we will fuse multiple data modalities to discover potential biomarkers for diagnostic outcome prediction and AD detection.

\vspace{-4pt}
\section{Conclusion}
\label{S6}
\vspace{-4pt}
In this paper, we investigate the 5-year cognitive progression of late-life depression (LLD) for the first time based on T1-weighted MRI and diagnosis information of long follow-up time.
This task is very challenging due to the 5-year time gap between MRI scan time and category labels (diagnosis information) and the heterogeneous relations between depression and cognitive impairment. 
We develop an HRL framework that can effectively fuse data-driven and handcrafted MRI features 
for longitudinal diagnostic discrimination in LLD. 
Experimental results on $294$ subjects with MRI acquired from two studies suggest the potential of our HRL in the automated diagnostic discrimination in LLD.
We further analyze the feature maps and find possible related ROIs that may contribute to the prediction of diagnostic cognitive change in LLD.
We also discuss the factors that limit deep learning models via  experiments and possible solutions to promote learning performance with limited data. 

\vspace{-4pt}
\bibliographystyle{model2-names.bst}
\biboptions{authoryear}
\bibliography{refs}

\end{document}

%% file: defs.tex
\def\x{{\bf x}}

\def\0{{\bf 0}}
\def\1{{\bf 1}}

\def\ie{{\em i.e.}}